\documentclass[aps,pra,twocolumn,showpacs,groupedaddress,amsmath,amssymb]{revtex4-1}

\usepackage{graphicx}
\usepackage{dcolumn} 
\usepackage{bm} 
\usepackage{hyperref}
\usepackage[caption=false]{subfig}

\hypersetup{
	colorlinks = true,
	urlcolor   = blue,
	linkcolor  = blue,
	citecolor  = blue
}

\DeclareMathOperator{\Tr}{Tr}

\DeclareMathOperator{\rank}{rank}

\renewcommand*{\phi}{\varphi}
\renewcommand*{\epsilon}{\varepsilon}
\renewcommand*{\le}{\leqslant}
\renewcommand*{\ge}{\geqslant}

\newcommand*{\prob}{\mathbb P}
\newcommand*{\argmax}{\mathop{\mathrm{argmax}}}
\newcommand*{\bra}[1]{\langle #1 |}
\newcommand*{\ket}[1]{| #1 \rangle}
\newcommand*{\bks}[2]{\langle #1 | #2 \rangle} 

\begin{document}
\preprint{APS/123-QED}
\title{Experimental Adaptive Quantum Tomography of Two-Qubit States}

\author{G.I.Struchalin$^1$}
	\email{glebx-f@mail.ru}
\author{I.A.Pogorelov$^1$}
\author{S.S.Straupe$^1$}
\author{K.S.Kravtsov$^{1,2}$}
\author{I.V.Radchenko$^{1,2}$}
\author{S.P.Kulik$^1$}
\affiliation{$^1$Faculty of Physics, M.\,V.\,Lomonosov Moscow State University, Moscow, Russia}
\affiliation{$^2$A.\,M.\,Prokhorov General Physics Institute RAS, Moscow, Russia}

\date{\today}

\begin{abstract}
We report an experimental realization of adaptive Bayesian quantum state tomography for two-qubit states. Our implementation is based on the adaptive experimental design strategy proposed in~\cite{Houlsby_PRA12} and provides an optimal measurement approach in terms of the information gain. We address the practical questions, which one faces in any experimental application: the influence of technical noise, and behavior of the tomographic algorithm for an easy to implement class of factorized measurements. In an experiment with polarization states of entangled photon pairs we observe a lower instrumental noise floor and superior reconstruction accuracy for nearly-pure states of the adaptive protocol compared to a non-adaptive. At the same time we show, that for the mixed states the restriction to factorized measurements results in no advantage for adaptive measurements, so general measurements have to be used.
\end{abstract}

\pacs{03.65.Wj, 03.67.-a, 02.50.Ng, 42.50.Dv}

\maketitle

\section{\label{sec:Intro}Introduction}

Quantum state tomography~(QST) is a procedure allowing to estimate an unknown state of a quantum system from the outcomes of measurements on an ensemble of identically prepared copies of this system. Since this primitive is essential for experimental quantum information studies, much effort was devoted to optimization of state tomography. An optimal tomographic protocol should demonstrate the best possible accuracy of reconstruction for a given number of performed measurements. It is clear, that \emph{adaptive} measurement strategies, where the choice of the next measurement depends on the previous outcomes may be advantageous~\cite{Freyberger_PRA00}. First implementations were carried out in a Bayesian framework and used precomputed decision trees to choose optimal measurement strategies~\cite{Wunderlich_PRA02}. Obviously, such an approach is applicable only for a very limited sample sizes. Recent advances in Bayesian methods, inspired by machine-learning applications, led to the development of fast algorithms allowing for an online determination of optimal measurements~\cite{Granade_NJP12, Houlsby_PRA12}. On the other hand, frequentist approach makes use of maximum likelihood estimation and suggests to determine the optimal measurements by minimizing the variance of locally unbiased estimators~\cite{Fujiwara_JPA06}. The latter ideas were implemented experimentally for a single parameter estimation~\cite{Takeuchi_PRL12, Stefanov_OptLett14}. Two recent experimental implementations~\cite{Steinberg_PRL13, Kravtsov_PRA13} have demonstrated the quadratic enhancement in the accuracy of reconstruction, measured as the infidelity with the ``true'' state, for adaptive estimation of qubit states. The experiment in~\cite{Steinberg_PRL13} relied on a single adaptive step: when half of the data are collected, the most likely state is estimated and the following measurements are performed in the eigenbasis of the estimated state. The work~\cite{Kravtsov_PRA13} used the fully adaptive Bayesian protocol based on maximizing the information gain for each measurement for a given Bayesian posterior distribution~\cite{Houlsby_PRA12}. Other utility functions besides information gain were considered for optimization, for example the infidelity-optimized Bayesian protocol for pure states was considered in~\cite{Hen_NJP15}. The Bayesian framework for state estimation may be advantageous for several reasons. First of all, the posterior distribution provides the most complete statistical description of the knowledge about the state inferred from the measurements, and never yields unphysical or unjustified estimates~\cite{BlumeKohout_NJP10}. Even more importantly Bayesian experimental design techniques offer rich possibilities for construction and optimization of adaptive tomographic protocols.

Adaptive tomography was previously implemented for single-qubit states only. Estimation of higher-dimensional states introduces new features, which should be addressed. For example, available measurements are usually restricted in a specific way, e.g. experimental system is split into two parts and projective measurements on each part can be implemented easily, while performing the most general measurement on the whole system is experimentally challenging~\cite{Steinberg_PRL10}. In this case the optimization of measurements in the adaptive protocol should be performed over some subset. Besides that, there are purely computational difficulties arising from the exponentially increasing dimensionality of the parameter space. In this work we address these issues and present the experimental realization of the fully adaptive Bayesian quantum state estimation protocol for two-qubit states. Our software is implemented for arbitrary dimensions and the experiment is performed for polarization state of entangled photon pairs. Two types of measurements are considered:
\begin{enumerate}
	\item general projective measurements,
	\item factorized measurements that can be represented as a tensor product of projective measurements on subsystems.
\end{enumerate}
Another issue in real experiments is the influence of the instrumental noise. We study the performance of adaptive tomography under noisy conditions, and find it advantageous over non-adaptive protocols. The paper is organized as follows: first, we review the inference algorithm and design of our experimental realization in Section~\ref{sec:Algorithm}; then in Section~\ref{sec:Exp}, we present experimental results and results of numerical simulations; finally, the conclusions are given in Section~\ref{sec:Conclusion}.

\section{Tomography algorithm\label{sec:Algorithm}}
Each tomographic protocol consists of two essential parts: a measurement strategy and an estimator. The measurement is characterized by a set of positive operator valued measures (POVM's) $\mathcal{M} = \{\mathbb M_a\}$ with index $\alpha \in \mathcal A$ numbering different configurations of the experimental apparatus. Having the measured system in the state $\rho$ and given the configuration $\alpha$, the probability of observing an outcome $\gamma$ is given via Born's rule:
\begin{equation}\label{eq:BornRule}
\prob(\gamma | \alpha, \rho) = \Tr(M_{\alpha\gamma} \rho),
\end{equation}
where $M_{\alpha\gamma} \in \mathbb M_\alpha$ are POVM elements. We will assume that each POVM is complete $\sum_\gamma M_{\alpha\gamma} = \mathbb{I}$. The estimator maps the set of all observed outcomes $\mathcal D_N = \{\gamma_n\}_{n=1}^N$ onto an estimation of the state $\hat \rho$.

\subsection{Bayesian approach}
Bayesian approach to quantum tomography deals with a probability distribution over the space of density matrices. First, a \emph{prior} distribution $p(\rho)$ is specified. Then, after data acquisition a \emph{posterior} probability $p(\rho | \mathcal D)$ is obtained via Bayes' rule:
\begin{equation}\label{eq:BayesRule}
p(\rho | \mathcal D) \propto \mathcal L(\mathcal D ; \rho) p(\rho),
\end{equation}
where $\mathcal L(\mathcal D ; \rho) = \prod_{n=1}^N \prob(\gamma_n|\alpha_n, \rho)$ is a \emph{likelihood} function, which incorporates a statistical model. It is equal to the probability of observing data $\mathcal D$ given the state $\rho$. The posterior distribution provides the most complete description of our knowledge about the quantum state, inferred from data $\mathcal D$~\cite{BlumeKohout_NJP10}.

The posterior distribution can be utilized for both statistical assessment and constructing a measurement strategy. Point estimates of desired quantities are found as expectations over the posterior distribution. Their error bars are given by variances. For example, we may obtain \emph{Bayesian mean estimate} (BME) of the state $\hat \rho = \mathbb E_{p(\rho | \mathcal D)} [\rho]$. The measure of its uncertainty is the distribution size $\bar d^2 = \mathbb E_{p(\rho | \mathcal D)} [d^2(\rho, \hat \rho)]$, where $d(\cdot,\cdot)$ is a distance between two states. In our analysis we used Bures metric $d_B$, which is closely related to the \emph{fidelity} $F(\rho, \hat \rho) = \Tr^2 \sqrt{\rho^{1/2} \hat \rho \rho^{1/2}}$~\cite{Jozsa_JMO94}: $d_B^2(\rho, \hat \rho) = 2-2 \sqrt{F(\rho, \hat\rho)} \approx 1-F(\rho, \hat\rho)$ for $1-F \ll 1$.

\paragraph*{Prior.\label{sec:Prior}}
For the purposes of quantum tomography the prior distribution over the state space should be non-informative or uniform. It is natural to identify uniformity with maximal symmetry. In the case of pure states such prior distribution is unique and it is determined by the requirement of unitary invariance~\cite{Jones_AnnPhys91}. Therefore the Haar measure is used. But for mixed states unitary invariance only implies the Haar measure over eigenvectors of density matrices, leaving the distribution in the simplex of eigenvalues $\lambda_i$ undetermined (each density matrix has a unit trace, so a set of eigenvalues forms a simplex $\{\lambda_i: \sum_i \lambda_i = 1\}$).

There are several approaches to specify a probability distribution of eigenvalues~\cite{Zyczkowski_JPA01}. The first one is based on a purification procedure: the $D$-dimensional system of interest is supplemented with a $K$-dimensional environment, and the composite $D \times K$ system is assumed to be in a pure state. The measure on the composite system is known (Haar measure) and after tracing out the environment, it induces a certain measure $d\mu_{DK}$ in the state space of the initial system. The second approach relies on the fact that each metric $d$ generates a measure $d\mu_d$: any ball with a fixed radius has the same measure. Interestingly, the measure $d\mu_{DK}$ induced by partial tracing for $D = K$ coincides with the measure induced by the Hilbert~-- Schmidt distance $d_{HS}^2(\rho, \hat{\rho}) = \Tr[(\rho - \hat{\rho})^2]$: $d\mu_{HS} = d\mu_{DD}$.

We shall consider the following prior distributions~\footnote{Strictly speaking in the formulas below one should add theta $\theta(\rho)$ and Dirac-delta $\delta(\Tr \rho -1)$ functions to ensure positive semi-definiteness and unit trace constraints on density matrices.}:
\begin{enumerate}
	\item induced by Bures distance
	\begin{equation}\label{eq:pBrho}
	p_{B}(\rho) \propto \prod_{i=1}^D \frac{1}{\sqrt{\lambda_i}}
	\prod_{i<j}^{D} \frac{1}{\lambda_i + \lambda_j},
	\end{equation}
	\item induced by Hilbert~-- Schmidt distance 	\begin{equation}\label{eq:pHSrho}
	p_{HS}(\rho) \propto 1,
	\end{equation}
	\item uniform in the eigenvalues simplex \begin{equation}\label{eq:pDeltarho}
	p_{\Delta}(\rho) \propto \prod_{i<j}^{D} \frac{1}{(\lambda_i - \lambda_j)^2}.
	\end{equation}
\end{enumerate}
After integrating over eigenvectors one obtains probability distributions $p(\lambda_i, \dots, \lambda_D)$ in the eigenvalues simplex. The expressions are formally similar to the right hand sides of~(\ref{eq:pBrho})~-- (\ref{eq:pDeltarho}) except the appearance of a \emph{geometric factor} $\prod_{i<j}^{D} (\lambda_i - \lambda_j)^2$. For instance,
\begin{equation}\label{eq:pHSlambda}
p_{HS}(\lambda_i, \dots, \lambda_D) \propto  1 \times \prod_{i<j}^{D} (\lambda_i - \lambda_j)^2.
\end{equation}
It is clear that $p_{\Delta}(\lambda_i, \dots, \lambda_D) \propto  1$ is indeed a uniform distribution in the eigenvalues simplex.

This geometric factor causes zero probability for the states with degenerate eigenvalues in both Hilbert~-- Schmidt and Bures distributions. At the same time, degenerate states are important, as they appear when a pure state $\ket{\psi}$ passes through a depolarizing channel $\ket{\psi}\bra{\psi} \to (1 - \epsilon) \ket{\psi}\bra{\psi} + \epsilon \mathbb I /D$, where $\mathbb I /D$ is a $D$-dimensional completely mixed state. Thus, the third, ``unnatural'', prior is introduced essentially to avoid such gaps at degenerate states. In appendix~\ref{sec:PriorInfluence} it is shown that the accuracy of a degenerate state tomography depends rather strongly on the prior choice and a potential explanation of this behavior is given.

\subsection{Adaptivity criterion\label{sec:AdaptivityCriterion}}
The Bayesian approach is a natural framework for constructing an adaptive measurement strategy. Indeed, the posterior can be updated as soon as a new outcome is observed, and given this new knowledge about the state one may find the next optimal measurement. There are several criteria for optimal experiment design known in literature (see Ref.~\cite{Murao_PRA12} for a review). One possible way is to choose an experimental configuration $\alpha$ that is expected to reduce Shannon entropy $\mathbb{H}$ of the posterior distribution the most. Usage of Shannon entropy as an objective function allows to reformulate this criterion as the following optimization procedure (which is more simple to work with in practice)~\cite{Houlsby_PRA12}:
\begin{equation}\label{eq:EntropyAdaptive}
\alpha = \argmax_{\alpha \in \mathcal{A}} \Bigl( \mathbb{H}[\prob(\gamma | \alpha, \mathcal{D})] - \mathbb{E}_{p(\rho | \mathcal{D})} \mathbb{H}[\prob(\gamma | \alpha, \rho)] \Bigr),
\end{equation}
where $\prob(\gamma | \alpha, \mathcal{D}) = \int \prob(\gamma | \alpha, \rho) p(\rho | \mathcal{D}) \, d\rho = \prob(\gamma | \alpha, \hat\rho)$ is the expected probability of outcome $\gamma$.
In our experimental realizations the maximization is performed among all POVM's $\mathbb{M}_\alpha$ consisting of projectors onto only \emph{factorized} states. On the other hand, \emph{general} projective measurements, without this constraint imposed, are investigated by means of numerical simulations.

Though the criterion~(\ref{eq:EntropyAdaptive}) seems complicated, one may provide intuition on how optimal measurements look like in the limit of a small distribution size. Expanding entropy $\mathbb{H}[\prob(\gamma | \alpha, \rho)]$ up to the second order in $\rho$ near the mean $\hat \rho = \mathbb E_{p(\rho | \mathcal D)} [\rho]$, we obtain that the entropy gain under maximization in~(\ref{eq:EntropyAdaptive}) approximately equals
\begin{equation}\label{eq:EntropyExpanded}
\sum_{\gamma} \frac{\mathbb{E}_{p(\rho | \mathcal{D})} \bigl(\prob(\gamma | \alpha, \rho) - \prob(\gamma | \alpha, \hat\rho) \bigr)^2}{\prob(\gamma | \alpha, \hat\rho)} \le \sum_{\gamma} \frac{\bar{d}_B^2}{\prob(\gamma | \alpha, \hat\rho)},
\end{equation}
where several inequalities for the trace distance $d_{tr}(\rho, \hat{\rho}) = \frac12 \Tr |\rho - \hat\rho |$ were used~\cite{NielsenChuang}: $\prob(\gamma | \alpha, \rho) - \prob(\gamma | \alpha, \hat\rho) \le d_{tr}(\rho, \hat{\rho})$ and $d_{tr} \le \sqrt{1-F} \le d_B$. The upper bound has the largest value when probabilities $\prob(\gamma | \alpha, \hat\rho)$ in the denominator are minimized. Therefore, expansion~(\ref{eq:EntropyExpanded}) suggests to project onto states that give low count rates~\footnote{High count rates will also appear, because all probabilities sum to unity $\sum_\gamma^\Gamma \prob (\gamma | \alpha, \hat \rho) = 1$. For example, if $\Gamma - 1$ probabilities achieve the smallest value, then the last one will have the highest value.}. In the general case the minimal probabilities are achieved in the eigenbasis of the current estimate $\hat\rho$. Measurements in the eigenbasis were exactly the adaptive strategy implemented in~\cite{Steinberg_PRL13}.

\subsection{Realization}
\paragraph*{Sampling.}
Normalization of the posterior distribution~(\ref{eq:BayesRule}) and averaging to obtain Bayesian mean estimates involves calculation of high-dimensional integrals. As the experiment progresses it becomes increasingly harder to perform on-line adaptive Bayesian inference (because the likelihood function consists of more terms). To solve this problem one has to use approximate inference usually based on Markov chain Monte Carlo (MCMC) methods. Here we briefly review the variant of sequential importance sampling (SIS) algorithm, proposed in~\cite{Houlsby_PRA12}.

In SIS one specifies a set of samples (or particles) $\{\rho_s\}_{s=1}^S$ with corresponding weights $\{w_s: \sum_s w_s = 1\}$. These samples approximate the posterior distribution as follows: $p(\rho | \mathcal D) \approx \sum_s w_s \delta(\rho - \rho_s)$. Thus, all integrals over the posterior are replaced by weighted sums over samples. Suppose after performing $n$ measurements, a new measurement in a configuration $\alpha_{n+1}$ yields the result $\gamma_{n+1}$. Then weights $w_s^{(n)}$ are updated according to the rule $w_s^{(n+1)} \propto w_s^{(n)} \prob(\gamma_{n+1} | \alpha_{n+1}, \rho_{s})$. Such an update of the weights is fast (requires $\mathcal O(1)$ operations), as it does not require recalculation of the full likelihood at every step (which has a cost of $\mathcal O(n)$). Therefore, SIS is very helpful for an adaptive experiment design.

At the same time, SIS has its own limitations: while the algorithm proceeds, the weights tend to collapse, i.e. all but a few of them become essentially zero. Consequently, the quality of the approximation drops down. To monitor the situation the effective sample size $S_{eff} = \bigl( \sum_{s=1}^{S} w_s^2 \bigr)^{-1}$ is calculated. If it is below a certain threshold then resampling is performed and the weights are equalized. A detailed description of the resampling procedure is provided in the appendix~\ref{sec:Resampling}.

At the start of the protocol samples are generated based on the chosen prior. For Hilbert~-- Schmidt $p_{HS}(\rho)$ and Bures $p_{B}(\rho)$ induced priors we utilized algorithms from~\cite{Zyczkowski_JMP11, Mezzadri_AMS07}, while the prior uniform in the eigenvalues simplex $p_\Delta(\rho)$ was constructed in two steps: first, samples are generated according to $p_{HS}(\rho)$ and, second, samples are moved by the resampling routine with the probability distribution~(\ref{eq:pDeltarho}).

\paragraph*{Block measurements.}
Achieving the best performance implies application of the adaptivity criterion after each successful measurement. However, this may be impractical due to the excessive burden of changing the measurement configuration. Instead, one can use block measurements by keeping the same configuration for $b$ consecutive measurement outcomes. Our study showed that having the block size $b = \mathcal O(N)$, where~$N$ is the total number of outcomes observed so far, does not noticeably impact the accuracy of the adaptive tomography; in particular we used $b = \lceil N/50 \rceil$ throughout the present work.

There is some evidence that a block size may be as large as $\mathcal O(N)$. Let us denote for simplicity $d^2_{true}(N) \equiv d^2_B(\hat \rho(N), \rho_0)$, which is the distance between the estimator $\hat \rho(N)$ and the true state $\rho_0$. It was shown in~\cite{Kravtsov_Arxiv13} that the measurement basis is optimal if it is aligned according to the true state with the error~$d^2_{align}$ smaller than~$d^2_{true}(N)$: $d^2_{align} = k \times d^2_{true}(N)$, $0 < k < 1$. After measuring the block of maximal size $b$, the basis ceases to be optimal: $d^2_{align} = d^2_{true}(N + b)$. With these two equations at hand we obtain $d^2_{true}(N + b) = k \times d^2_{true}(N)$. Assuming the behavior $d^2_{true}(N) = c N^a$ (see Sec.~\ref{sec:Exp}), finally, we find that block size scales linearly with $N$: $b = (k^{1/a} - 1) N = \mathcal O(N)$.


Interestingly, our simulations shows that the amount of information extracted during one block $b \propto N$ (the entropy gain under maximization in~(\ref{eq:EntropyAdaptive}) times the block size $b$) holds approximately constant, while the tomography proceeds. This can be used to construct a rule for selecting a block size on the fly: the block lasts until the given amount of information is not gathered. Another valid approach, where changes of the measurement configuration are based on its deviation from the optimal, was used in~\cite{Hen_NJP15}. The authors have shown that on average the number of configuration adjustments~$y$ grows logarithmically with~$N$, and therefore the block size $b \approx dN/dy \propto N$ again scales linearly with $N$.

\section{Implementation\label{sec:Exp}}
\subsection{Experiment}
\begin{figure}[th]
	\includegraphics[width=\linewidth]{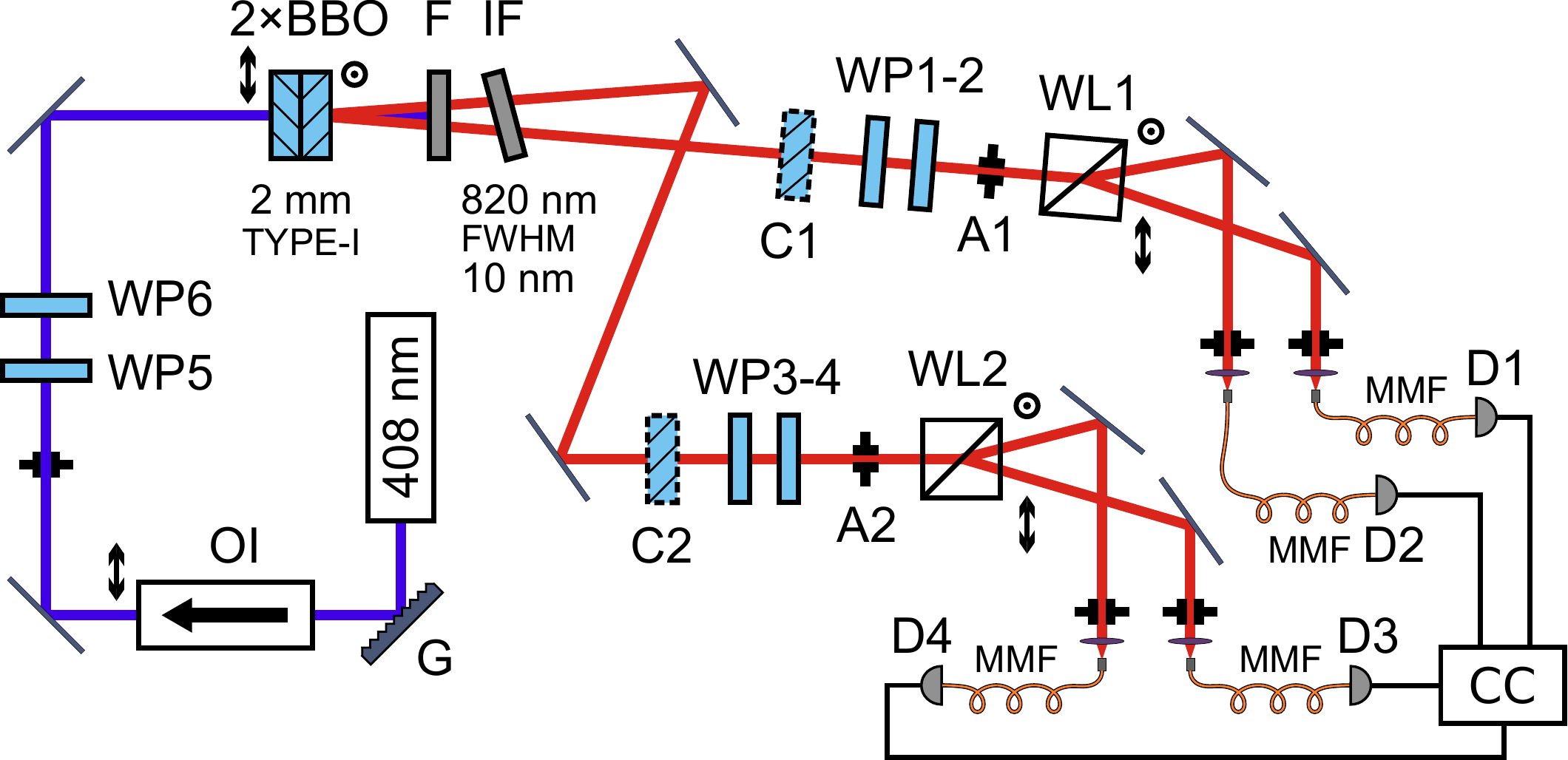}
	\caption{Experimental setup. The entangled photon pairs are generated in two BBO crystals with perpendicular axes, pumped by linearly polarized light from a 408~nm diode laser. Projective measurements are performed separately for each photon of the pair by means of half and quarter wave plates followed by Wollaston prisms.}
	\label{img:Setup}
\end{figure}

The described protocols are implemented experimentally using polarization degrees of freedom of photon pairs produced by spontaneous parametric down-conversion (SPDC). The sketch of the experimental setup is shown in Fig.~\ref{img:Setup}. State preparation is performed by pumping two non-linear BBO crystals with linearly polarized light. Littrow configuration~\cite{Hawthorn_RSI01} external cavity diode laser (408~nm) is used as a pump. The Faraday isolator~OI serves for both blocking parasitic light, reflected backwards, and preparing initial linear polarization. This polarization is then rotated by means of two wave plates~WP5-6. The crystals are cut for a slightly non-collinear frequency-degenerate type-I phase-matching. Such configuration allows us to prepare polarization-entangled states of variable degree of entanglement and purity~\cite{Kwiat_PRA99}. The purity of the generated state can be altered by changing the size of the collecting irises A1-2 or by inserting the birefringent compensators~C1-2 with high wavelength dependent phase shifts after the crystals~\cite{Kwiat_OptExpr09, Straupe_JETP10}. A degree of polarization entanglement is controlled by rotating the linear polarization of the pump.

The two components of generated photon pairs propagate through similar channels, which consist of wave plates~WP1-4 (zero order half and quarter wave plate in each channel) mounted on a computer-controlled rotation mounts and Wollaston prisms~WL1-2. Finally, they are coupled to multi-mode fibers connected to single photon counting modules (SPCM's)~D1-4. This setup allows us to make any factorized 4-dimensional projective measurement (i.e. realize projectors onto $|\psi\rangle=|\psi_1\rangle \otimes |\psi_2\rangle$, where $|\psi_{1,2}\rangle$ are the polarization states of photons in channels 1 and 2).

\begin{figure*}[th]
	\subfloat[Factorized state.]
	{	
		\includegraphics[width=0.49\linewidth]{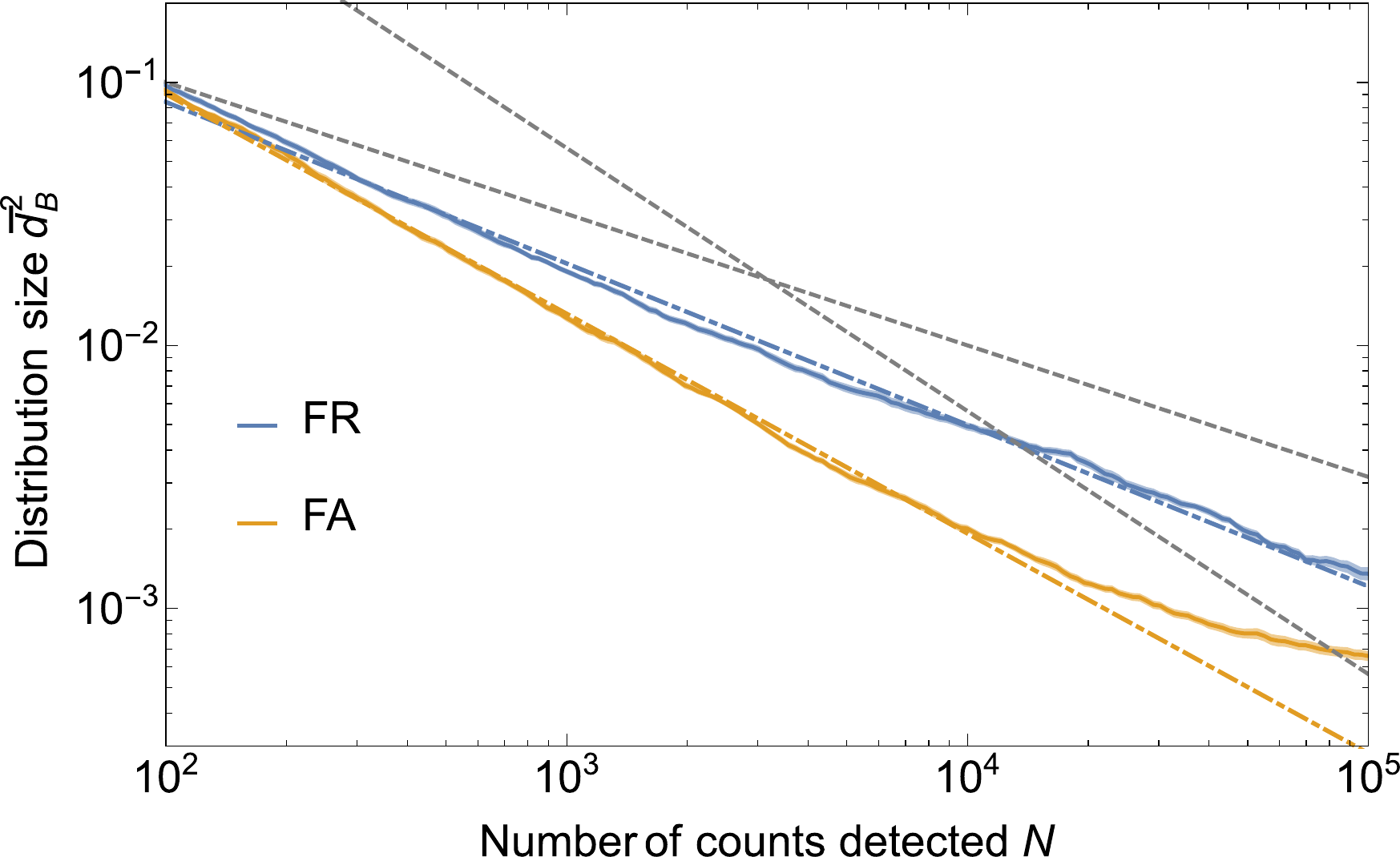}
		\label{img:HH}
	}
	\subfloat[Bell state.]
	{
		\includegraphics[width=0.49\linewidth]{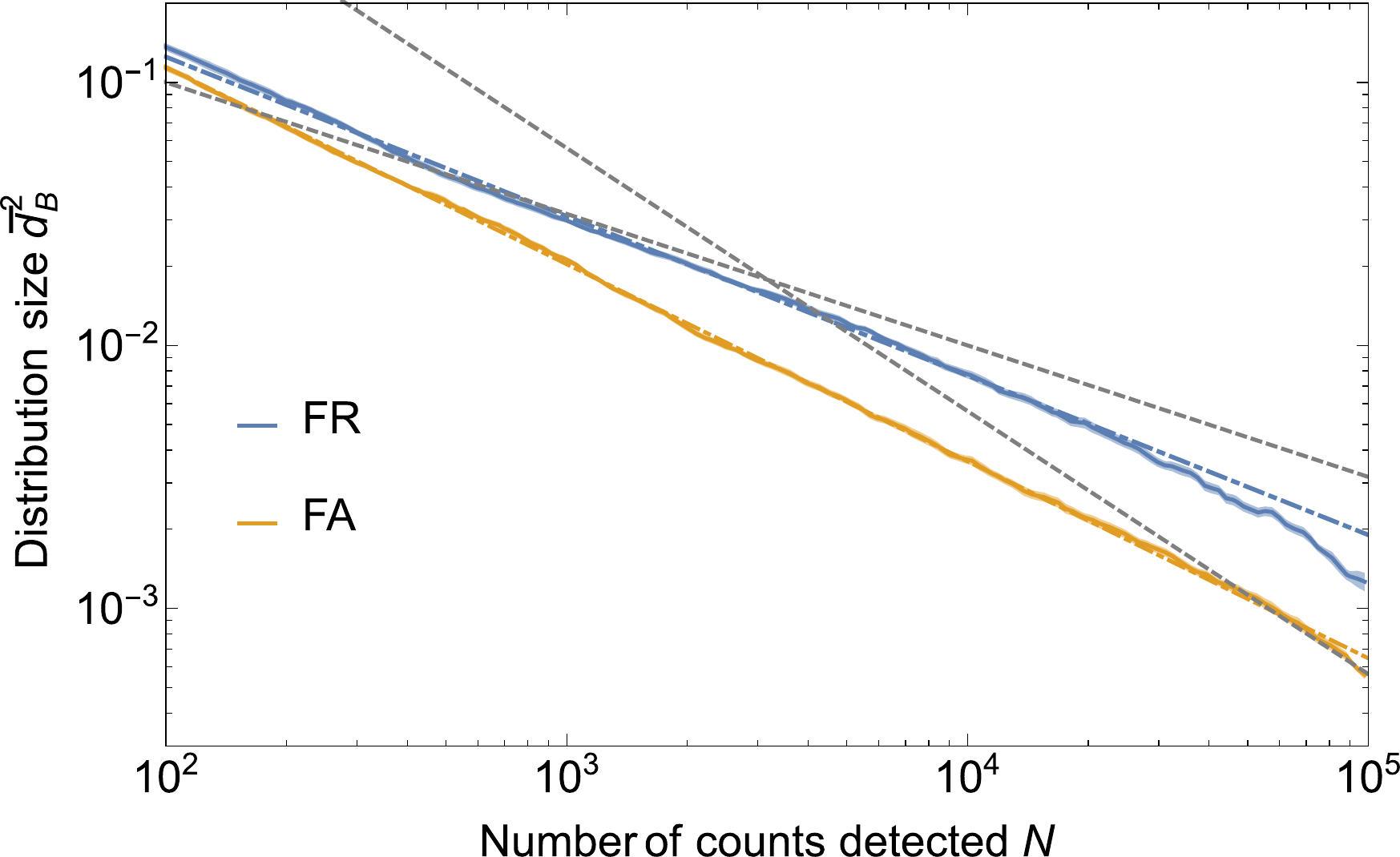}
		\label{img:Bell}
	}
	\caption{Experimental results. The experimentally obtained dependencies of the posterior distribution size $\bar d_B^2$ on the overall number of detected pairs for the cases of a separable (a) and maximally entangled (b) polarization state. The dot-dashed lines are power law fits to the data, and the dashed gray lines are the dependencies $\propto N^{-1/2}$ and $\propto N^{-1}$ shown here and on consecutive plots for comparison. Legend: $FR$ stands for factorized random measurements and $FA$~-- for factorized adaptive.}
	\label{img:ExpConvergence}
\end{figure*}

We have compared the performance of the adaptive measurement strategy with the random measurements strategy that is known to be optimal in a non-adaptive case~\cite[Theorem 3.1]{Holevo_book_1982}. A comparison of protocols was carried universally within the Bayesian framework. The Bayesian mean state $\hat \rho = \mathbb E_{p(\rho | \mathcal D}[\rho]$ was used as an estimate of the true state. A natural figure of merit for any estimation technique is the distance $d_B^2(\hat\rho, \rho_0)$ between the estimation $\hat\rho$ and the true state $\rho_0$. However, in a practical tomography the true state is usually unknown and the tomography provides the best estimation at hand: e.g. one may perform a very long series of measurements and use the final result as the ``true'' state. Its accuracy, though, is still limited by the experimental imperfections. Therefore, instead of using distance to the true state $d_B^2(\hat\rho, \rho_0)$ as a measure of accuracy, we turned to its assessment~-- the posterior distribution size $\bar d_B^2 = \mathbb E_{p(\rho | \mathcal D)} [d_B^2(\rho , \hat \rho)]$. In appendix~\ref{sec:DistrSize} it is shown that the distribution size $\bar d_B^2$ is in fact connected with $d_B^2(\hat\rho, \rho_0)$ and thus provides a reasonable estimation.

Fig.~\ref{img:ExpConvergence} shows the dependence of the posterior distribution size~$\bar d_B^2(N)$ on the number~$N$ of the detected pairs for the factorized random (FR) and factorized adaptive (FA) projective measurements. The data are a result of averaging over many (typically 15) single full runs of the tomography protocol. These results clearly demonstrate an advantage of the adaptive strategy over the random one for both factorized (Fig.~\ref{img:HH}) and maximally entangled (Fig.~\ref{img:Bell}) state tomography. The results of fitting the data with a power law in the form $cN^a$ are presented in Table~\ref{tab:ExpFits}. The adaptive protocol clearly demonstrates an improved distribution size scaling in the Bures metric, although it is not exactly the expected $N^{-1}$ dependence. We attribute this reduction in performance to the two main reasons: the inaccuracies in waveplates positioning (on the order of $0.2^\circ$) and deviations of the measured states from truly pure. The former was previously observed (see Supplementary information of~\cite{Steinberg_PRL13}) and is known to reduce the performance of the adaptive protocol. While the latter, studied below in the simulations section, is due to the degraded behavior of the restricted factorized adaptive protocol for mixed states compared to the ideal adaptive tomography; thus, the observed reduction of the convergence rate is also expected. There is an apparent saturation of the $FA$ protocol curve for a factorized state (Fig.~\ref{img:HH}). The most probable explanation is that a state under the tomography has happened to cause a plateau in the dependence~$\bar d_B^2(N)$ (see Appendix~\ref{sec:PriorInfluence}).


\begin{table}[th]
	\caption{Power law fits of the distribution size dependence on the number of detected pairs $\bar d_B^2(N)$ obtained in experiment for factorized random (FR) and factorized adaptive (FA) measurements. Digits in parentheses are uncertainties in accord with the fit.}
	\label{tab:ExpFits}
	\begin{ruledtabular}
		\begin{tabular}{ccc}
			State & FR & FA \\
			\hline
			Factorized state&
			$1.4(2) \times N^{-0.61(4)}$&
			$4.2(3) \times N^{-0.83(5)}$
			\rule{0pt}{11pt}\\
			Bell state&
			$2.0(4) \times N^{-0.60(6)}$&
			$3.6(0) \times N^{-0.74(9)}$
		\end{tabular}
	\end{ruledtabular}
\end{table}

\begin{figure}[bh]
	\includegraphics[width=\linewidth]{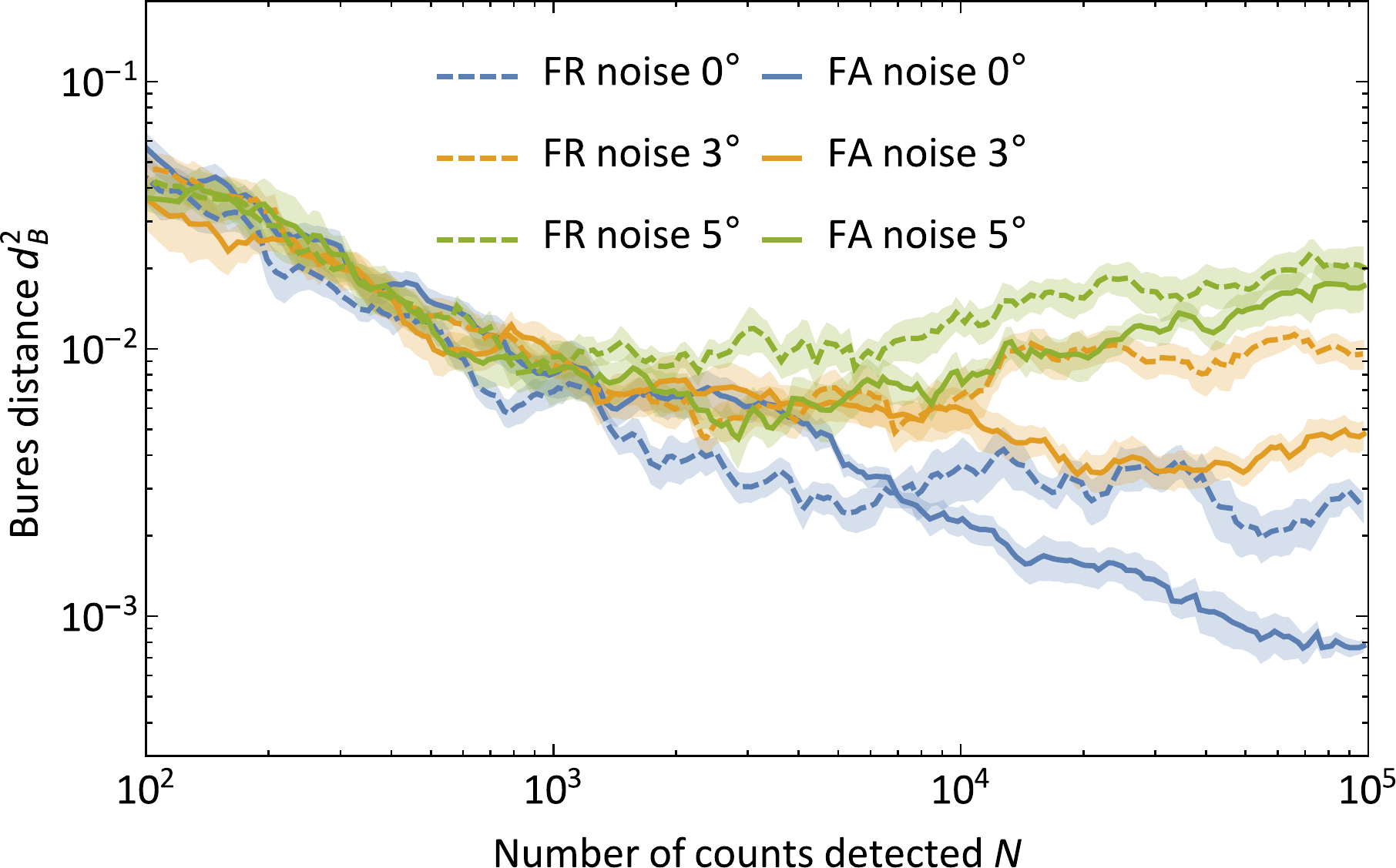}
	\caption{Experimental tomography in the presence of instrumental errors. The dependence of the \emph{spread} of experimental results (see text for definition) on the overall number of registered events. Dashed lines correspond to random protocol, while solid ones~-- to adaptive tomography. Color encodes different amounts of additional instrumental noise in the experimental run: no additional noise (blue), $3^\circ$ (dark yellow) and $5^\circ$ (green) of additional waveplates uncertainty. The shaded areas correspond to one standard deviation for the ensemble of~10 different tomography runs.}
	\label{img:Noisy}
\end{figure}

We have experimentally observed that the adaptive tomography of two qubits is less sensitive to instrumental errors than a non-adaptive one. Previously, the same was shown for a single qubit tomography by means of numerical simulations~\cite{Steinberg_PRL13}.
The following noise model was used: each time the wave plates WP1-4 are rotated, their positions are intentionally set to $\theta_i + \delta \theta_i$ rather than to the exact angles $\theta_i$ used by the estimation algorithm. Here $\delta \theta_i$ is a random variable uniformly distributed in the range $\left[-\Delta \theta, \Delta \theta\right]$. Again, due to the absence of knowledge about the true state, the quantity~$d_{spr}^2$, which we will refer to as \emph{spread} over different runs of tomography, was used. Let $\hat\rho_k(N)$ be the Bayesian mean estimate in the $k$-th experimental run. Then the spread $d_{spr}^2(N) = \frac1K \sum_{k=1}^K d_B^2 \bigl(\hat\rho_k(N), \sigma(N) \bigr)$, where $\sigma(N) = \frac1K \sum_{k=1}^K \hat\rho_k(N)$ is the mean of estimates at $N$-th step over $K$ runs. The spread reflects reproducibility of tomography results from run to run.

The experimentally obtained dependence of the spread on the number of detected pairs $d_{spr}^2(N)$ for $K=10$ runs of the tomography protocol for the Bell state is presented in Fig.~\ref{img:Noisy} where we compare the behavior for different noise amplitudes $\Delta \theta = \{0^\circ, 3^\circ, 5^\circ\}$. Zero amplitude means no additional ``software'' noise is added, and the noise floor is determined by the intrinsic errors in positioning introduced by the wave plates rotators ($\approx 0.2^\circ$). For moderate statistics $N \lesssim 10^3$ there is no visible difference between adaptive and random measurements. But when the dependencies saturate for larger $N$, a noise floor for the adaptive protocol is lower. Therefore, the attainable accuracy of the tomography is higher if one uses adaptive techniques, given the same imperfect measurement apparatus. As the amplitude $\Delta \theta$ increases, the noise floors for adaptive and random protocols tend to converge. In our case the difference becomes negligible for $\Delta \theta = 5^\circ$. It is worth to mention that even for $N = 10^5$ there is no visible saturation for the adaptive protocol without additional noise ($\Delta \theta = 0^\circ$), while the random protocol reaches its noise floor at $N \approx 10^4$.

Reduced sensitivity to errors in wave plates positioning for the adaptive tomography can be intuitively understood in the following way. If the exact probabilities of a tomographically complete set of measurements are known, than one can easily reconstruct the true state. In real world applications probabilities can only be determined with some finite accuracy. The higher is the accuracy the better is the assessment of the true state. Therefore the goal is to provide the best estimation of probabilities. Suppose we perform a tomography of a state $\rho_0$, and POVM elements $M_{\alpha\gamma} = M_\gamma (\theta_i)$ are parameterized by parameters $\theta_i$ (wave plates angles in our case). The adaptive tomography tends to select parameters $\theta_i$, that give low probabilities of outcomes $p_A(\theta_i)$ (see sec.~\ref{sec:AdaptivityCriterion}). Therefore, $p_A(\theta_i)$ take extremal values, and error in probability estimation $\delta p_{A}$ depends quadratically on the uncertainties in parameters $\delta\theta_i$: $\delta p_A = \mathcal O (\delta \theta_i^2)$. This is not the case for non-adaptive protocols, unless the POVM is aligned with the true state~\cite{Steinberg_PRL13}. For non-adaptive tomography typically $\delta p_R = \mathcal O (\delta \theta_i) \gg \delta p_A$.

\subsection{Simulations\label{sec:ResSim}}
The chosen experimental approach allows \emph{factorized} projective measurements only, which we refer to $F$ class measurements. Projective measurements of a general form ($G$ class), without factorization constraints, could also be implemented, although with a significantly larger experimental efforts~\cite{Steinberg_PRL10}. Since $G$ class is much wider it is reasonable to expect higher accuracy of the state estimation. In this section we focus on a question whether an extension $F \to G$ dominates over adaptivity or not. Several results for pure states  have been previously known from the literature. It was shown~\cite{Steinberg_PRL10} that non-adaptive measurements in mutually unbiased bases (which belong to $G$ class because they require projections onto entangled states) can improve precision for \emph{some} states. On average, general measurements are not worse than factorized ones (symbolically one can write $GR \ge FR$, where ``larger'' means more accurate). Though, the situation is different for adaptive protocols~\cite{Houlsby_PRA12}: on average, factorized adaptive measurements outperform general non-adaptive ($FA > GR$).
We have performed numerical simulations of four protocols with different types of measurements: factorized random ($FR$), factorized adaptive ($FA$), general random ($GR$), and general adaptive ($GA$). Our goal was to compare the average performance of Bayesian tomography for those four types.


\begin{figure*}
	\subfloat[]
	{
		\includegraphics[width=0.49\linewidth]{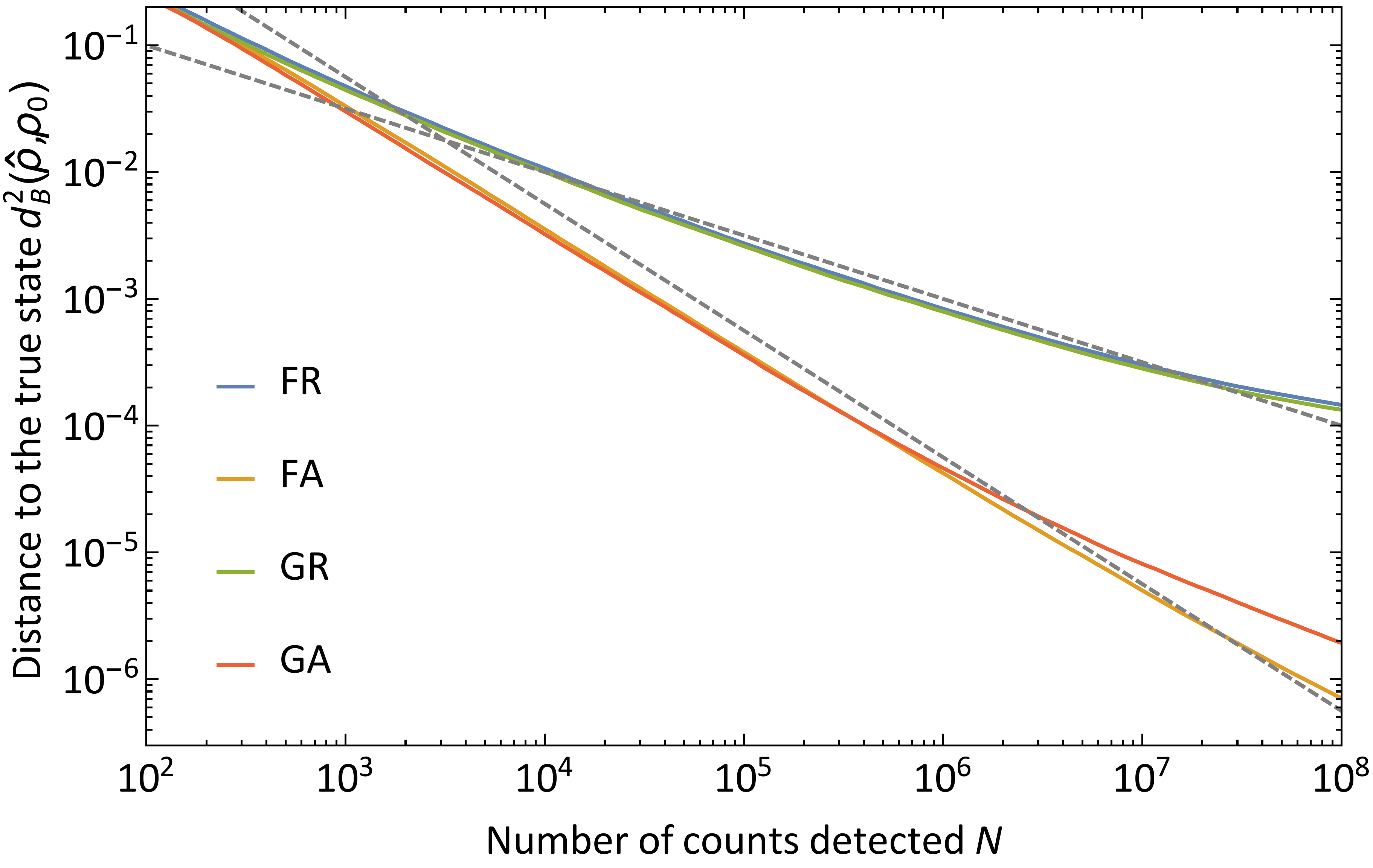}
		\label{img:PureAverage}
	}
	\subfloat[]
	{
		\includegraphics[width=0.49\linewidth]{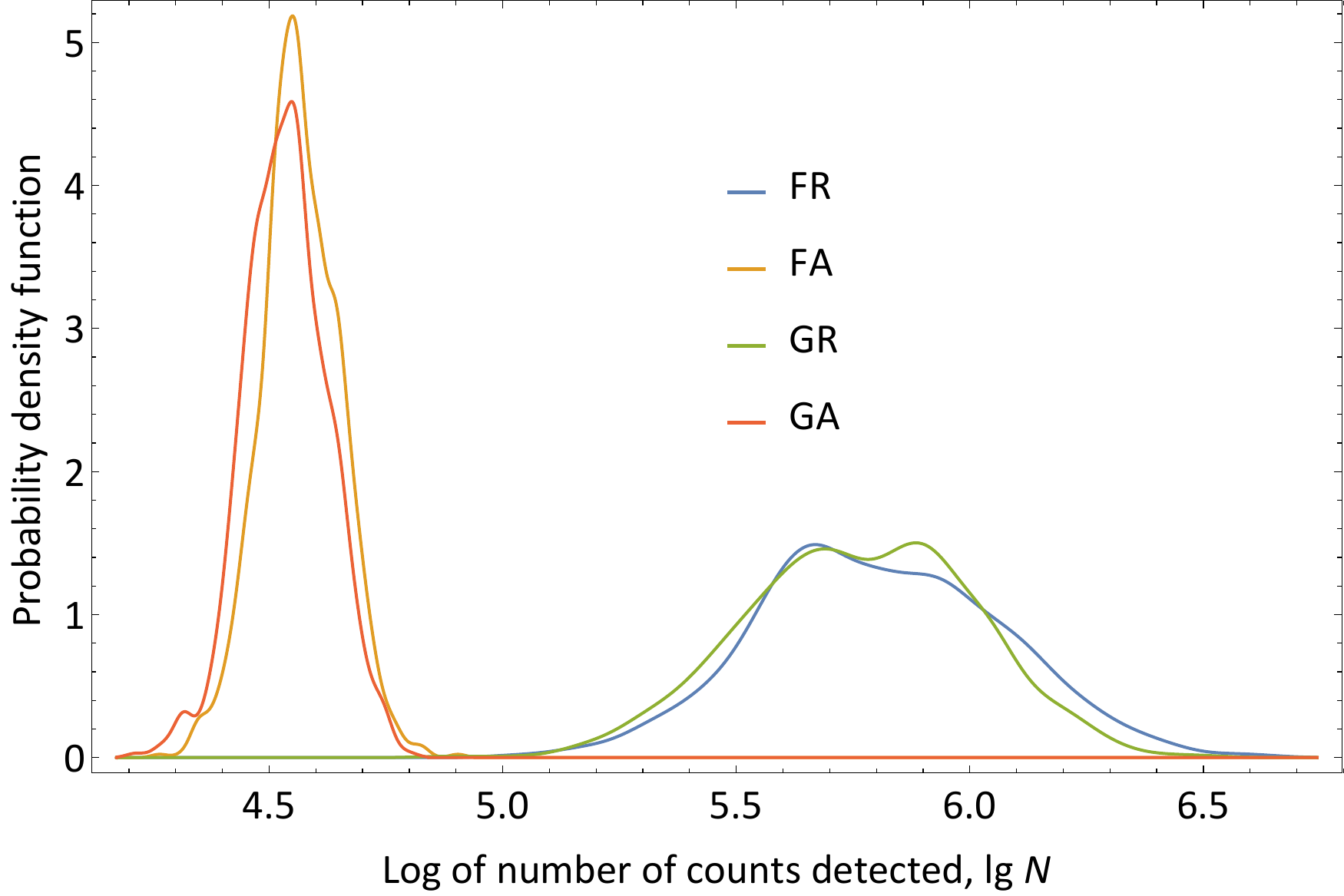}
		\label{img:PureHistogram}
	}
	\caption{The results of numerical simulations for pure states. The dependence of distance to the true state averaged over the posterior distribution on the total number of registered events $N$ (\ref{img:PureAverage}), and histograms for the values of $N$ required to reach a fixed accuracy $d_B^2 = 10^{-3}$ (\ref{img:PureHistogram}) are shown. Legend: $F$ stands for factorized measurements, $G$~-- for general, $R$~-- for random, $A$~-- for adaptive.}
	\label{img:Pure}
\end{figure*}

Let us start with the pure states. For each protocol we have averaged over 1000 runs with different pure true states, uniformly distributed with respect to the Haar measure. The simulation results are shown in Fig.~\ref{img:Pure}. We have tried various priors, defined according to eqs. (\ref{eq:pBrho})--(\ref{eq:pDeltarho}), and found, that the prior choice only changes the accuracy by a multiplicative constant, thus preserving a relationship between different protocol performance (see Appendix~\ref{sec:PriorInfluence} for details). Here we present the results for the prior uniform in the eigenvalues simplex $p_\Delta(\rho)$. As a figure of merit the distance $d_B^2(\hat\rho(N), \rho_0)$ between a current estimate $\hat\rho(N)$ and the true state $\rho_0$ is used (Fig.~\ref{img:PureAverage}). It is also instructive to show histograms for the values of $N$ for which a fixed accuracy $d_B^2 = 10^{-3}$ is reached in different runs (Fig.~\ref{img:PureHistogram}). The results indicate that the determining factor is adaptivity, because the performance is independent of the measurement class $F$ or $G$. There is an obvious enhancement in the estimation accuracy for adaptive protocols. Power law fits of the dependencies $d_B^2(\hat\rho(N), \rho_0)$ of the form $cN^a$ give \emph{convergence rates} $a=-0.958 \pm 0.008$ for adaptive protocols and $a=-0.588 \pm 0.011$ for the random ones, which is close to the theoretically expected scalings $N^{-1}$ and $N^{-1/2}$ respectively. To summarize, for pure states we have a relation $GA = FA > GR = FR$~-- on average there is no advantage in using general measurements.


\begin{figure*}
	\subfloat[]
	{
		\includegraphics[width=0.49\linewidth]{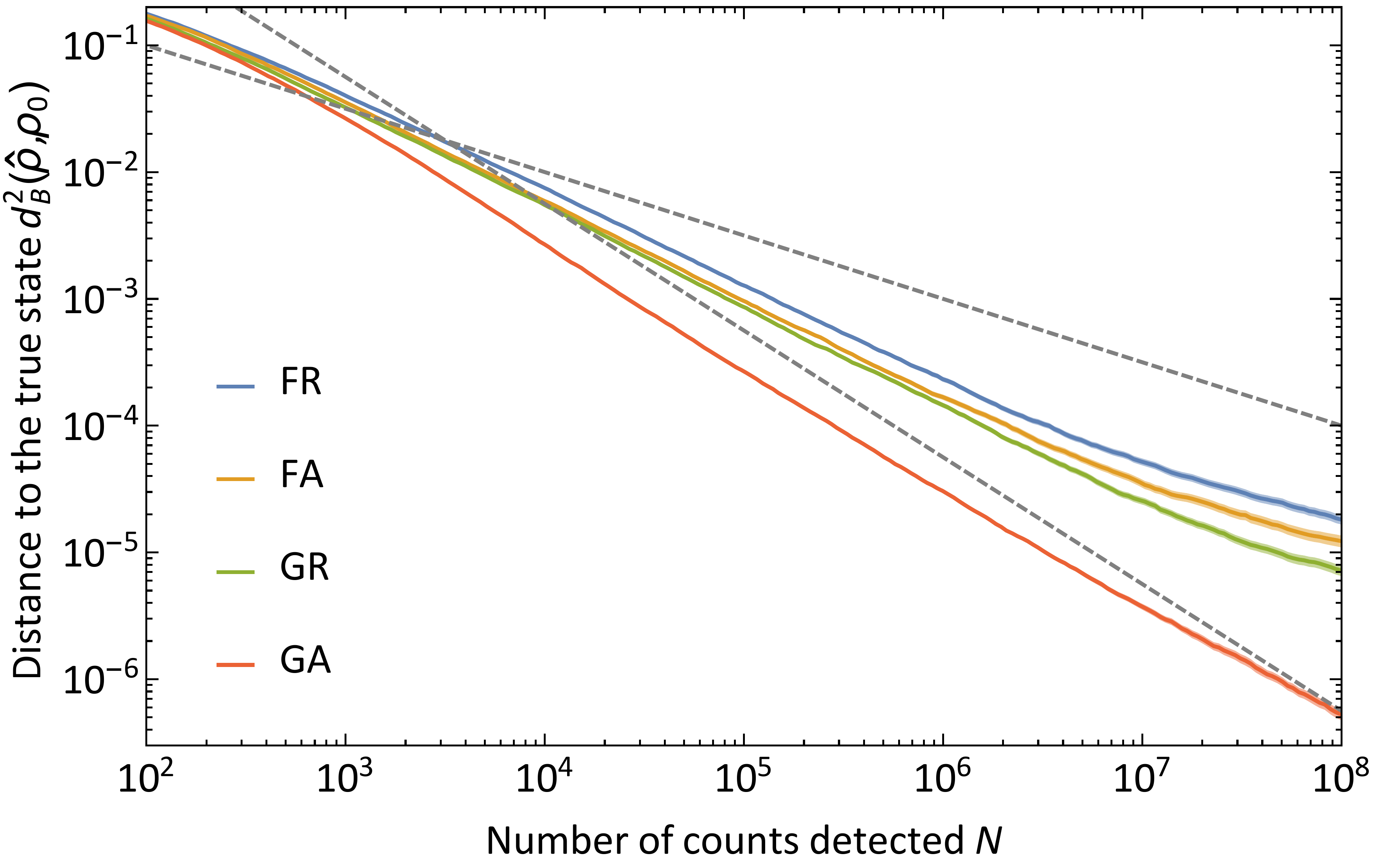}
		\label{img:BuresAverage}
	}
	\subfloat[]
	{
		\includegraphics[width=0.49\linewidth]{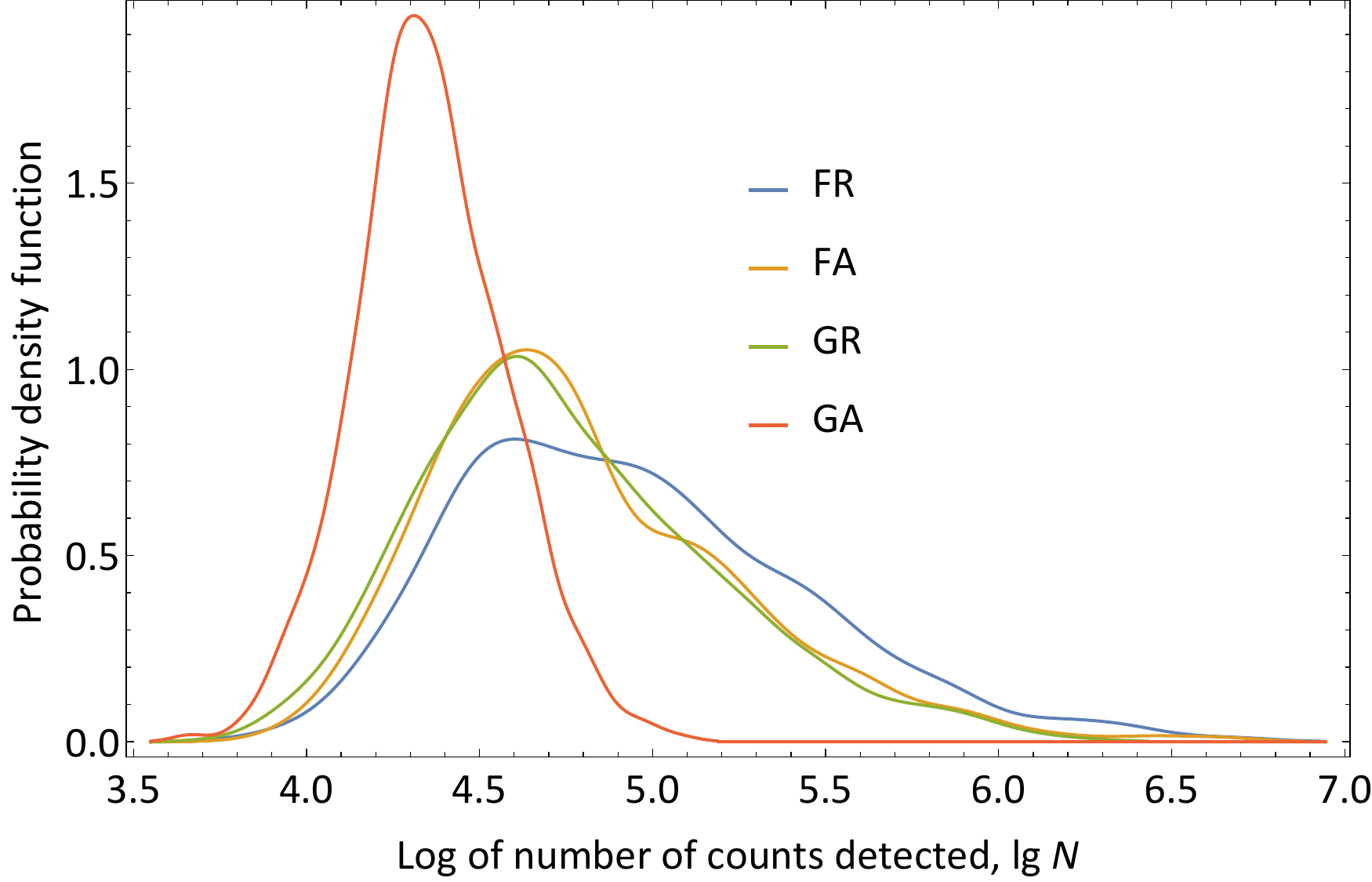}
		\label{img:BuresHistogram}
	}
	\caption{The results of numerical simulations for mixed states picked at random from the uniform distribution with respect to the Bures-induced measure. The dependence of a distance to the true state averaged over the posterior distribution on the total number of registered events $N$ (\ref{img:PureAverage}), and histograms for the values of $N$ required to reach a fixed accuracy $d_B^2 = 10^{-3}$ (\ref{img:PureHistogram}) are shown.}
	\label{img:Bures}
\end{figure*}

As genuine pure states are experimentally unfeasible, it is important to study how the protocol performs for the general form mixed states. The results of simulations averaged over mixed states are shown in Fig.~\ref{img:Bures}. The dependencies represent the average behavior over 1000 true states, randomly distributed with respect to the Bures-induced measure. The results are different from those for the pure states: $GA > FA = GR \gtrsim FR$. Only $GA$ protocol provides solid enhancement with $N^{-1}$ scaling ($a = -1.00 \pm 0.02$) for non-pure states, while other three protocols behave nearly similarly and perform at $N^{-3/4}$ level ($a = -0.77 \pm 0.03$). Previously, the average convergence $N^{-3/4}$ was predicted for non-adaptive mixed state tomography of qubits~\cite{Bagan_PRA04}. The accuracy of the $GA$ protocol remains unaffected as expected, but the performance of the $FA$ protocol is worse for mixed states as compared to pure ones. This is rather surprising because in the single qubit tomography it is known that mixed states are in some sense easier to estimate than pure~\cite{Steinberg_PRL13}.
Therefore, naively extending this result to the high-dimensional tomography, one may expect that accuracy will at least not be reduced for mixed states. However, this is not the case for the $FA$ protocol. At the same time the accuracy of non-adaptive protocols for mixed states is improved.

\begin{figure}
	\includegraphics[width=\linewidth]{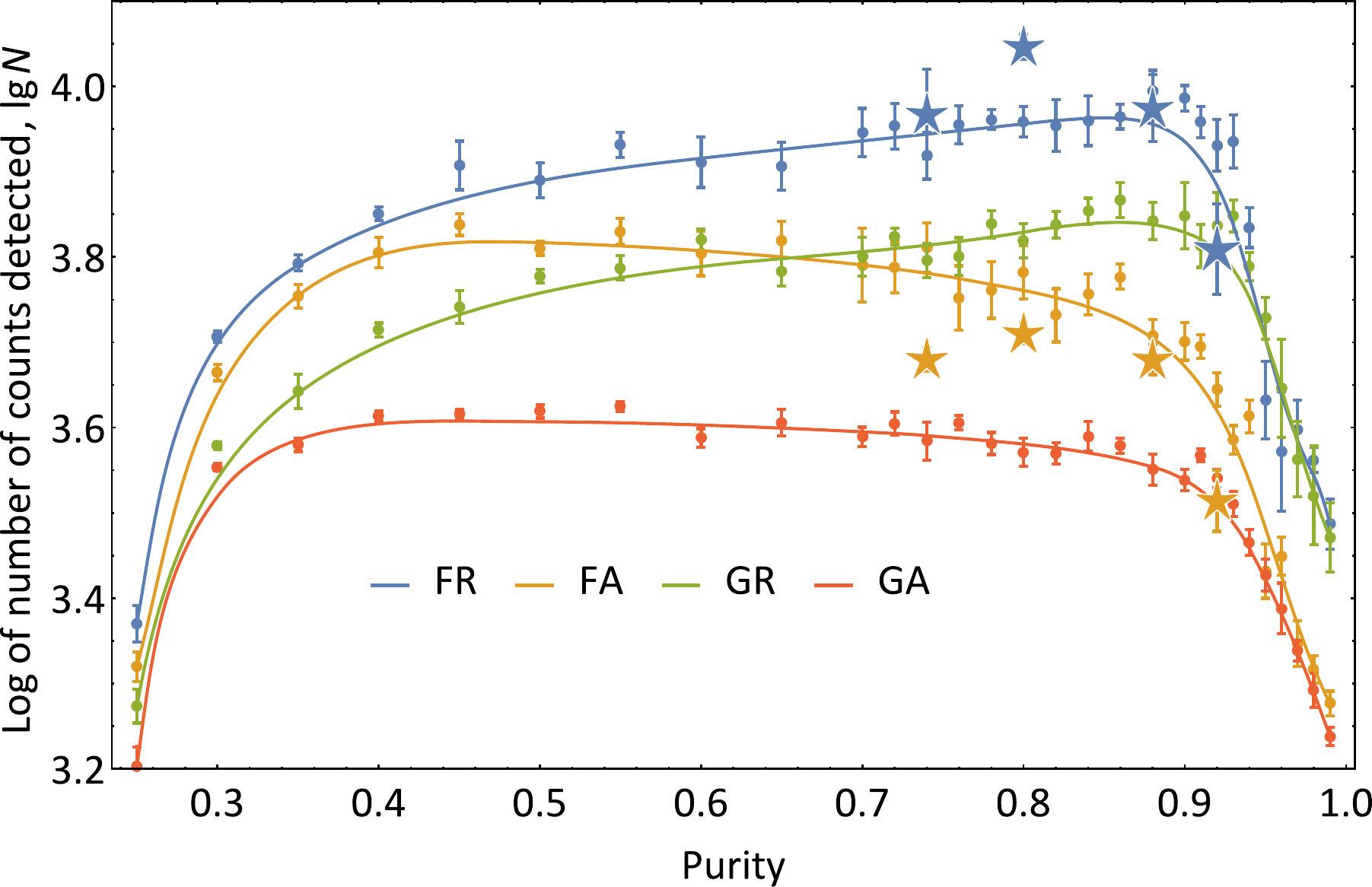}
	\caption{Tomography performance for states of different purity. The value of $N$ required to reach the specified precision, measured by the distance to the true state of $\bar d^2_B(N) = d^2_{err} = 5 \times 10^{-3}$, averaged within the given purity layer is shown. The stars correspond to experimental points, while dots are results of numerical simulation. Error bars show a standard deviation for averaged values of~$N$. The solid lines are guides to the eye.}
	\label{img:ReachVsPurity}
\end{figure}

From continuity considerations it is clear that for states pure enough, the pure state tomography behavior $GA = FA > GR = FR$ should still hold. To understand the transition between the two, we have performed averaging over several layers of states with different purity $\Tr \rho^2 = \mathrm{const}$. To construct each layer we effectively cut 10 appropriate true states from the Bures distribution. Then the distribution size dependence~$\bar d^2_B(N)$ is averaged over these states. Value of $N$ required to reach the specified precision, measured by the distance to true state of $\bar d^2_B(N) = d^2_{err} = 5 \times 10^{-3}$, is plotted against the purity in Fig.~\ref{img:ReachVsPurity}. For $FA$ and $FR$ protocols we also plot the experimentally obtained points. They are in reasonable agreement with the simulated data. For purities above $p_{pure} \approx 0.94$ the advantage of general measurements is negligible, and results for pure states hold ($GA = FA > GR = FR$). For purities below $p_{pure}$ general measurements are preferable for either adaptive ($GA > FA$) or non-adaptive protocols ($GR > FR$). On the other hand, the adaptivity is still dominating the accuracy ($FA > GR$) for purities above $p_{adapt} \approx 0.7$. However, the boundaries~$p_{adapt}$ and~$p_{pure}$ themselves depend on the selected error level~$d^2_{err}$. For lower error levels $p_{adapt}$ shifts towards smaller values, due to the different scaling for adaptive and non-adaptive protocols. On the contrary, $p_{pure}$ approaches unity, because nearly-pure states become tomographically distinguishable from truly pure for smaller~$d^2_{err}$ (for example, $p_{pure} \approx 0.98$ for $d^2_{err} = 10^{-3}$).

\section{Conclusion\label{sec:Conclusion}}

We have demonstrated an experimental realization of a fully adaptive Bayesian state estimation protocol for two-qubit states. Even with the restriction of the experimentally attractive factorized measurements performed separately on each qubit, the protocol shows significant improvement of the estimation accuracy compared to the optimal non-adaptive protocol. This improvement is most significant for pure and nearly-pure states. Achieving the advantage for mixed states, though, requires implementation of the joint entangling measurements on both qubits. This is somewhat unexpected, because mixed states are known to be easier to estimate with non-adaptive strategies~\cite{Steinberg_PRL13}. However, the most interesting area for quantum information applications~-- nearly-pure states~-- still shows very good performance, so the advantage of the general measurements probably does not justify the experimental efforts for their realization. Thus, the factorized adaptive protocol seems to be the optimal choice, considering the trade-off between accuracy and challenging implementation.

A natural question is whether it is possible to scale the protocol to higher dimensions? The straightforward answer will be~-- probably not too far, since the computation complexity grows exponentially with the dimensionality of the Hilbert space of the reconstructed system. This is, however, the case for any tomographic protocol, so the only extension here is the search for an optimal measurement, which is by itself not more demanding than the estimation of the posterior. All the computations, performed for the experiment presented here, were done online during the experimental run with a standard desktop computer, so the required computational power for two-qubits (15-dimensional parameter space) is reasonable. Some modifications may be introduced as well to allow running the algorithm for several qubits with tolerable performance. For example, the time-consuming resampling procedure, may be simplified by relaxing the requirement for exact sampling from the likelihood function, and substituted by some approximation, such as the Liu-West resampling algorithm~\cite{Liu_West_2001}, which only preserves the first and the second moment of the sampled distribution, as it was done for Hamiltonian learning in~\cite{Granade_NJP12}. This will push the tractable dimension boundary a little bit higher, however `the curse of dimensionality' is inevitable, and one should not expect that the full tomography of a several dozen qubit state will be feasible. Other avenues of research may be pursued here, such as restricting the tomography to some physically relevant subspace, e.g. matrix-product states~\cite{Cramer_Nature10}, or using compressed sensing techniques for low-rank states~\cite{Gross_PRL10}. It will be interesting to investigate the advantages of self-learning techniques, like those presented here, offered in these settings. In general, we expect to see new interesting applications for machine-learning and Bayesian methods in quantum state and process tomography and in other related areas.

\begin{acknowledgments}
The work was partially supported by  the RFBR grants no. 14-02-00705, 14-02-00749, 14-02-00765, and European Union Seventh Framework
Programme under Grant Agreement No. 308803 (project BRISQ2).
\end{acknowledgments}

\appendix

\section{Prior influence\label{sec:PriorInfluence}}
Though tomography accuracy, measured as a posterior average Bures distance to the true state, averaged over some ensemble of states, varies by a (small) multiplicative constant for different priors, a pointwise behavior can change dramatically. The reason is that a neighborhood of a true state may include states, which are rare or exceptional in the given prior, i.e. have low probability. Among the considered priors~(\ref{eq:pBrho})--(\ref{eq:pDeltarho}), Bures and Hilbert~-- Schmidt distributions assign zero probability to states with degenerate eigenvalues (see eq.~(\ref{eq:pHSlambda})). This behavior is related to the geometry of the density matrix space~\cite{Zyczkovsky_book_2006}. One can cancel out the geometric factor, obtaining, for example, a uniform in the eigenvalues simplex prior~$p_\Delta(\lambda) \propto 1$, which assigns non-zero probability for all states.

\begin{figure}
	\includegraphics[width=\linewidth]{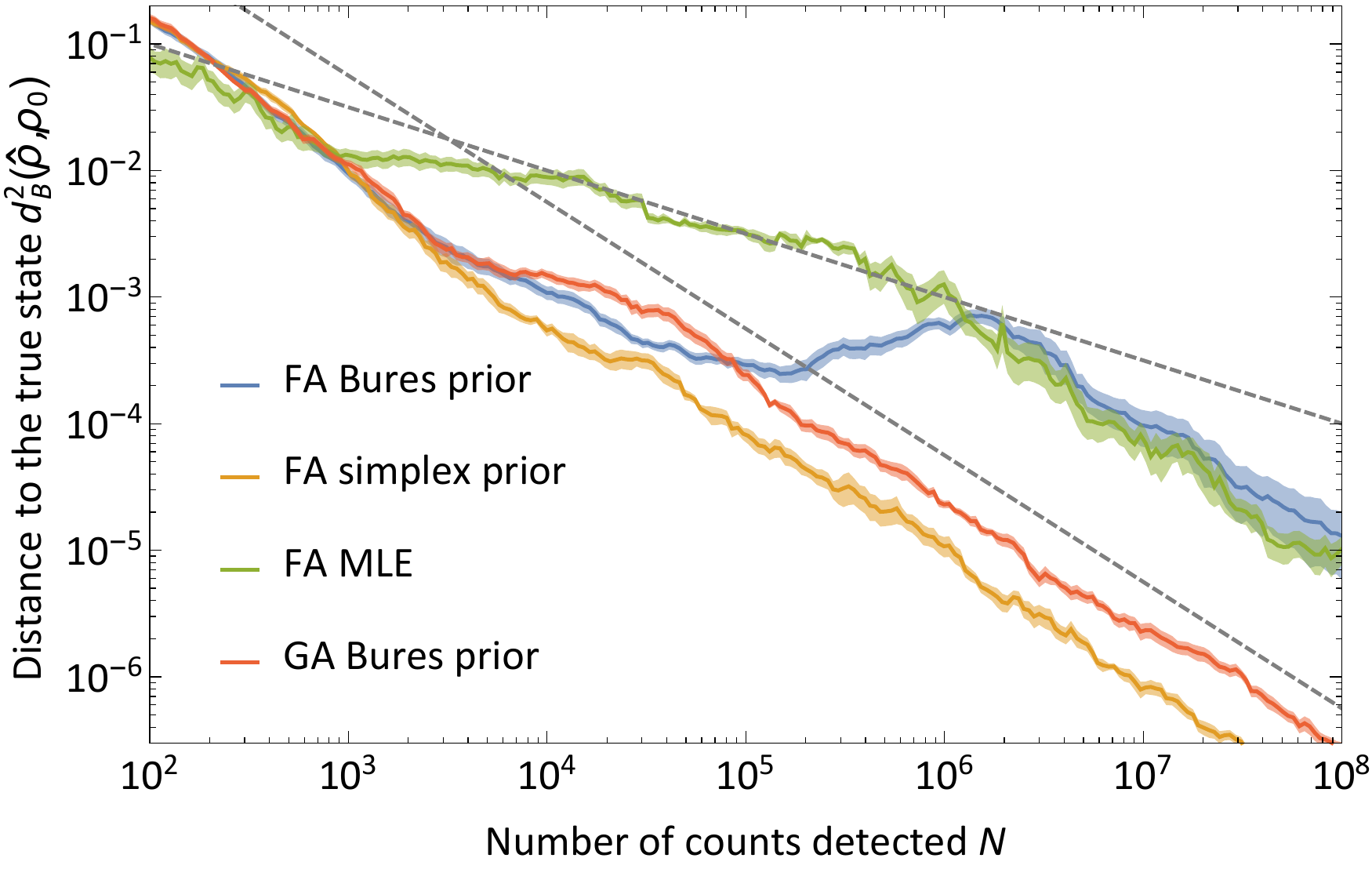}
	\caption{Influence of the prior distribution. The dependence of the average distance to the true state $d_B^2(\hat \rho(N), \rho_0)$ on the number of registered events $N$ for the $FA$ protocol with Bures (blue line) and uniform in simplex (yellow line) prior. For the $GA$ protocol the problem with convergence is less visible (red line). The distance of the maximum-likelihood estimate to the true state (green line) is shown for comparison. The shaded areas correspond to one standard deviation for the ensemble of~10 different tomography runs.}
	\label{img:Bend}
\end{figure}

The dependence of the average distance to the true state $d_B^2(\hat \rho(N), \rho_0)$ on $N$ is compared for the Bures and simplex priors for the $FA$ protocol in Fig.~\ref{img:Bend}. The Hilbert~-- Schmidt prior gives rise to a similar behavior as the Bures one. The true state is chosen to have a diagonal density matrix with a triply degenerate eigenvalue: $\lambda_i = \{0.9925, 0.0025, 0.0025, 0.0025\}$ (this state has a purity of $\approx 0.985$). For the $FA$ protocol with the Bures prior the scaling is close to $N^{-1}$ in the beginning but then a plateau occurs. When a sufficient amount of data is obtained the normal behavior $N^{-1}$ restores, however, the prefactor~$c$ becomes larger by two orders of magnitude. As a consequence of the plateau existence, accuracy does not increase during a long interval $2 \times 10^5 \lesssim N \lesssim 4 \times 10^6$, and it looks like tomography is useless in this range. The simplex prior recovers the situation: the scaling $N^{-1}$ and the initial prefactor remain the same during the whole tomography run.

As a reference we have utilized the maximum likelihood estimator (MLE; numerical algorithm is the one from~\cite{Lavor_QIC14}), which does not involve any priors inherent to the Bayesian inference (though we used an optimal measurement set found by the $FA$ protocol with the simplex prior). As compared to the BME with the Bures prior, the MLE demonstrates a $N^{-1/2}$ scaling instead of the plateau, nevertheless the accuracy is worse. Asymptotically at large~$N$ the MLE behaves similarly to the BME within error bars. Altogether, the MLE is not better than BME.

We have presented the results for an extreme case of a triply degenerate density matrix. For the states which lie farther from degenerate ones, the plateau becomes shorter. A position of the plateau $d_{plt}^2$ is proportional to a distance between the true state $\rho_0$ and the border of the space of physical density matrices, formed by rank-3 states: $d_{plt}^2 \propto \min_{\rank \rho = 3} d_B^2(\rho_0, \rho) \approx \lambda_{min}$ for $\lambda_{min} \ll 1$, where $\lambda_{min}$ is the minimal eigenvalue of the true state. It is worth to mention that the plateau shape is protocol dependent: for $GA$ it is significantly shorter.

For critical applications, to achieve the best performance, we suggest construction of an informative prior from physical considerations, e.g. based on a particular decoherence model for a given system. For a recent recipe of constructing useful informative priors, see~\cite{Granade_Arxiv15}.

\section{Distribution size\label{sec:DistrSize}}
\begin{figure}
	\includegraphics[width=\linewidth]{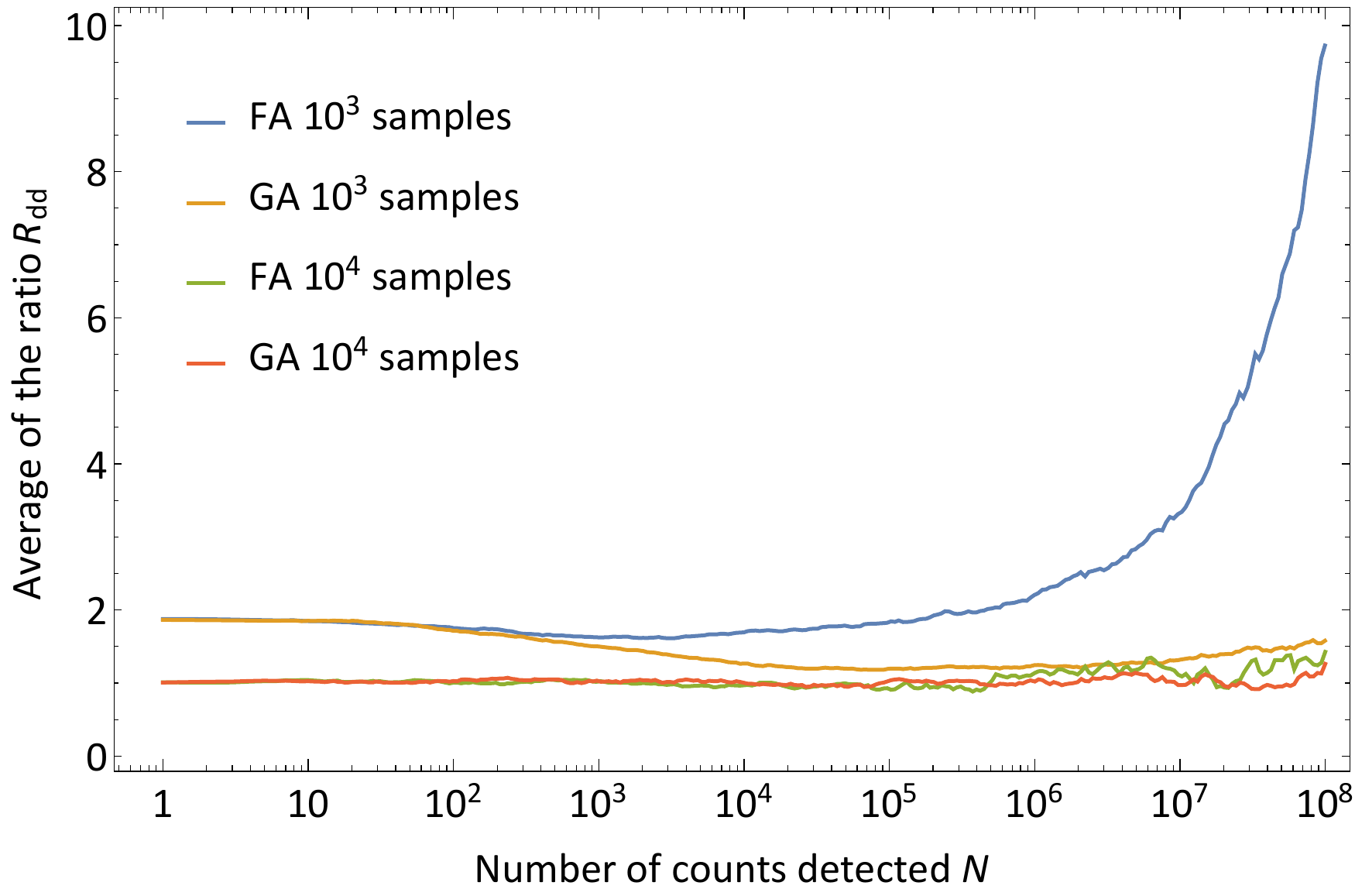}
	\caption{The average ratio $R_{dd}$ of the Bures distance to the true state $d_B^2(\hat \rho(N), \rho_0)$ and the posterior size $\bar d_B^2(N)$ in Bures metric. Simulated results for $GA$ and $FA$ protocols and different number of samples used in the sequential Monte-Carlo calculation are shown. It is clear, that for a sufficient number of samples (which is lower for the $GA$ protocol) the posterior distribution size is a reasonable estimate of the tomographic accuracy.}
	\label{img:RatioDD}
\end{figure}

In order to verify that the distribution size~$\bar d_B^2$ can be considered as an appropriate estimation of the tomography accuracy $d_B^2(\hat \rho, \rho_0)$ we studied how the ratio $R_{dd} = d_B^2(\hat \rho, \rho_0) / \bar d_B^2$ of the distance to the true state and the distribution size evolves with~$N$ in numerical simulations. The ratios~$R_{dd}(N)$, found for single tomography runs with different true states, randomly chosen from the Bures distribution, have been averaged over 1000 runs. The behavior of the averaged ratio $R_{dd}$ for $10^3$ and $10^4$ samples used in the Monte-Carlo calculations is compared in Fig.~\ref{img:RatioDD}. For $10^3$ samples the ratio holds almost constant, not larger than 2, till $N \approx 10^5$ so the accuracy $d_B^2(\hat \rho, \rho_0)$ in an experiment can be estimated by means of the distribution size $\bar d_B^2$. For larger $N$ the sampling technique should be refined by increasing the number of samples to guarantee that the ratio remains constant. Interestingly, the behavior of $R_{dd}(N)$ is protocol dependent: for the $GA$ protocol $10^3$ samples are sufficient within the whole interval $1 \le N \le 10^8$, while for the $FA$ protocol this is not the case, because the ratio starts to increase heavily from $N \approx 10^5$.

The distance to the true state $d_B^2(\hat \rho, \rho_0)$ is a coarse estimate of the tomography accuracy, because it gives only one number to characterize a region in the state space, where the true state is likely situated; it does not provide any information about the shape of the region. A more rigorous approach in Bayesian tomography is to use region estimates, e.g. posterior covariance ellipsoids~\cite{Ferrie_NJP14}, to specify the credible region.

\section{Resampling\label{sec:Resampling}}
The resampling procedure is used to redraw samples $\{\rho_s\}_{s=1}^S$ from the posterior distribution $p(\rho | \mathcal D)$ when the effective sample size $S_{eff} = \bigl( \sum_{s=1}^{S} w_s^2 \bigr)^{-1}$ is below a certain threshold $S_{eff} < S_{thrs}$. In our inference we typically use $S = 1000$ samples and a threshold of $S_{thrs} = 0.1 \times S$. The resampling algorithm consists of several steps:
\begin{enumerate}
	\item pick a particle $\rho_s$ from the sample distribution with a probability $w_s$, repeat $S$ times. At this point particles with tiny weights are eliminated and the ones with high weights are duplicated;
	\item equalize all weights $w_s := 1/S$;
	\item move each particle according to the Metropolis~-- Hastings (MH) algorithm~\cite{Hastings_Bio70} to sample the posterior distribution correctly.
\end{enumerate}
The MH procedure requires a random step rule $\rho_0 \to \rho$ (also called a proposal distribution $Q(\rho | \rho_0)$), according to which a candidate particle $\rho'$ is generated. This particle is then accepted or rejected based on the \emph{acceptance ratio}
\begin{equation}\label{eq:Acceptance}
	r = \frac{p(\rho' | \mathcal D)}{p(\rho_0 | \mathcal D)} \times \frac{Q(\rho | \rho')}{Q(\rho' | \rho)}.
\end{equation}
Here the posterior probability $p(\rho | \mathcal D)$ has to be calculated using a full-likelihood function~(\ref{eq:BayesRule}). This is a time consuming task and therefore a trade-off between resampling accuracy and calculation speed exists. In our implementation 50 MH accept/reject iterations were found to be sufficient.

The random walk of the MH algorithm relies on the purification procedure~\cite{BlumeKohout_NJP10}. Every density matrix $\rho_0$ can be purified in a space of extended dimension $D \times D$, i.e. $\rho_0 = \Tr_2 \ket{\psi_0}\bra{\psi_0}$, where $\Tr_2$ is a partial trace over auxiliary system and $\ket{\psi_0}$ is a pure state in the extended system. `Cholesky-like' matrix decomposition $\rho_0 = A A^\dagger$ for the density matrix is useful to perform the purification in practice. Indeed, one can easily check, that $\ket{\psi_0} = \sum_{i,j} A_{ij} \ket{ij}$ with $\ket{ij}$ being the standard computational basis. Once the purification $\ket{\psi_0}$ is found, the random step $\mathcal W$, followed by partial tracing, is applied
\begin{equation}\label{eq:RandomStep}
	\ket{\psi} = \mathcal W (\ket{\psi_0}) = a \ket{\psi_0} + b \cdot \frac{\ket{g} - \ket{\psi_0}\bks{\psi_0}{g}}{\|\ket{g} - \ket{\psi_0}\bks{\psi_0}{g}\|},
\end{equation}
where $a, b$ are the parameters controlling the step size ($a^2 + b^2 = 1$) and $\ket{g} \sim \mathcal N(0, \mathbb I)$ is a vector with real and imaginary parts of its elements being i.\,i.\,d.\ random variables with normal distribution $\mathcal N(0, 1)$. We choose a Gaussian step size in Bures metric by setting $a = 1-d^2/2$ with $d \sim \mathcal N(0, \sigma)$. To achieve a nearly constant fraction of accepted steps the standard deviation $\sigma$ was set proportional to the posterior distribution size $\bar d_B$.

Let us multiply the random operator $\mathcal W$ by an arbitrary unitary matrix $U$. Then the generated probability distribution does not depend on the order of $\mathcal W$ and $U$ in the product (however, the commutator $[\mathcal W, U] \ne 0$):
\begin{equation}\label{eq:RandomStepCommute}
	U \mathcal W(\ket{\psi_0}) \sim \mathcal W(U \ket{\psi_0}).
\end{equation}
This can be proved by a substitution of the explicit form of $\mathcal W$~(\ref{eq:RandomStep}) into~(\ref{eq:RandomStepCommute}). The left-hand and the right-hand sides will be equal up to the substitution $\ket{g} \to U^\dagger \ket{g}$. However, $U^\dagger \ket{g} \sim \mathcal N(0, \mathbb I)$ is again from the same ensemble of Gaussian vectors as $\ket{g}$ is. Let us consider $U$ in~(\ref{eq:RandomStepCommute}) as an arbitrary unitary operator with a fixed point: $U\ket{\psi_0} = \ket{\psi_0}$. Than we find that the distribution of vectors $\ket{\psi} = \mathcal W (\ket{\psi_0})$ is invariant under such unitary transformations. Thus, the expression~(\ref{eq:RandomStep}) generates a ``spherical'' neighborhood of the center $\ket{\psi_0}$ with a radial probability distribution determined by $a$. This property is desired for the MH algorithm because it allows to explore the space more isotropically, reducing the required number of iterations.

In our case the random step procedure is based on purification procedure and hence induces a uniform Hilbert~-- Schmidt probability distribution $p_{HS}(\rho) \propto 1$. As a result, the second multiplier in~(\ref{eq:Acceptance}) vanishes: $Q(\rho | \rho') / Q(\rho' | \rho) = 1$. Indeed, if one sets a uniform probability distribution $p_{MH}(\rho) \equiv 1$ as a desired probability in the MH routine, ($p(\rho | \mathcal D) \propto p_{MH}(\rho)$), then $r = 1$, i.e. all steps are accepted. So the algorithm will converge to the distribution induced by a random walk $p_{HS}(\rho)$, which coincides with the desired distribution $p_{MH}(\rho)$.

For other types of a random walk the second multiplier in~(\ref{eq:Acceptance}) may not be equal to 1. Nevertheless, we believe that one can force the multiplier to unity and use a random step that inherently generates, for example, the Bures-uniform distribution. The development of such random step rule would be an eligible task.

\suppressfloats
\bibliography{ref_base,mainNotes}

\end{document}